\documentclass{ws-procs975x65}
\def\lsim{\lower.5ex\hbox{$\; \buildrel < \over \sim \;$}}
\def\gsim{\lower.5ex\hbox{$\; \buildrel > \over \sim \;$}}
\def\be{\begin{equation}}
\def\ee{\end{equation}}

\begin{document}

\title{Spectral Properties of a Two Component and Two Temperature Advective Flow }

\author{SAMIR MANDAL}
\address{Centre for Space Physics, Chalantika 43, Garia Station Rd., 
Garia, Kolkata, 700084\\ e-mail: space\_phys@vsnl.com}

\author{SANDIP K. CHAKRABARTI\footnote{\uppercase{A}lso at \uppercase{C}entre for 
\uppercase{S}pace \uppercase{P}hysics, \uppercase{C}halantika 43, \uppercase{G}aria 
\uppercase{S}tation \uppercase{R}d., \uppercase{K}olkata 700098}}
\address{S.N. Bose National Centre for Basic Sciences,\\
JD-Block, Salt Lake, Kolkata 700098\\
E-mail: chakraba@bose.res.in}

\maketitle

\abstracts {Low angular momentum accretion flows very often have centrifugal pressure 
supported standing shock waves which can accelerate flow particles. The 
accelerated particles in turn emit synchrotron radiation in presence of 
magnetic fields. Efficient cooling of the electrons reduces its temperature 
in comparison to the protons. In this paper, we assume two temperature flows to
explore this property of shocks and present an example of the emitted radiation spectrum.}

\noindent To be Published in the proceedings of the 10th Marcel Grossman Meeting (World Scientific Co., Singapore),
Ed. R. Ruffini et al.

\section{Introduction}

Chakrabarti (1989) showed that the centrifugal barrier in a low angular momentum
accretion flow can produce stable shocks. Chakrabarti and 
Wiita (1992) and Chakrabarti and Titarchuk (1985) showed that shocks 
could play a major role in determining the spectrum of the emitted radiation.
Particularly important is that, the post-shock region, which is the 
repository of hot electrons, can easily inverse Comptonize photons from 
a Keplerian disk located in the pre-shock region and the power-law component of the
flow may be formed easily without taking resort to any hypothetical electron cloud. 
However, photons can also be generated by thermal or magnetic bremsstrahlung
and they can also be comptonized by the hot electroons.
Just as cosmic rays may be accelerated by shocks in supernovae explosions (Bell 1978ab; 1981),
the {\it standing shocks} in accretion disks, through which much of the accreting matters pass
before entering into a black hole, or forming a jet, should be  
important to energize matter. The resulting power-law electrons would also
inverse Comptonize along with maxwell-Bolzmann electrons. In the present paper, we are interested 
to identify the signatures of this energetic matter in the spectra of black holes.

\section{Solution Procedure}

For a given case, we fix the outer boundary at a large distance (say, $10^6 r_g$) 
and supply matter (both electrons and protons) with the same temperature 
(say, $T_p=T_e=10^6K$). We calculate density and temperature at any point
by solving the energy equations separately for electrons and protons.
We compute the radiation emitted by the flow through bremsstrahlung
and synchrotron radiation. These low energy photons are then 
inverse Comptonized by the hot electrons in the flow. We followed the
procedures presented in Chakrabarti and Titarchuk (1995) and Titarchuk and
Lyubarskij (1994) for computing the Comptonized spectrum due to Maxwell-Boltzmann
electrons and power-law electrons respectively. At the end, we add the 
contributions to get the net photon emissions from the flow. The geometry 
of the flow is chosen to be conical. The angle $\Theta$ subtended 
by the flow with the z-axis is chosen to be a parameter. For simplicity, 
we also choose the shock location $X_s$ and the compression ratio $R$ 
as free parameters. The shock of compression ratio $R$ causes 
the formation of power-law electrons of slope $p=(R+2)/(R-1)$ (Bell, 1978ab). This
power-law electrons produce a power-law of the synchrotron
emission with slope $q=(1-p)/2$ (Longair 1981). The power-law electrons
have energy minimum at $E_{min}=3\theta$, where, $\theta$ is electron 
temperature, and energy maximum at $E_{max}$ obtained self-consistently
by conserving the number of power-law electrons and by computing the 
number of scattering that the electrons undergo inside the disk before 
they escape. (Here $E$ is the bulk Lorentz factor of the electrons.) 
In a realistic flow, not all of the incoming matter is expected 
to pass through the accretion shock, and we assume that the percentage 
of electrons $\zeta$ having power-law index to be a free parameter.

\begin{figure}[t]
\vskip -5.0cm
\centerline{\epsfxsize=4.0in\epsfbox{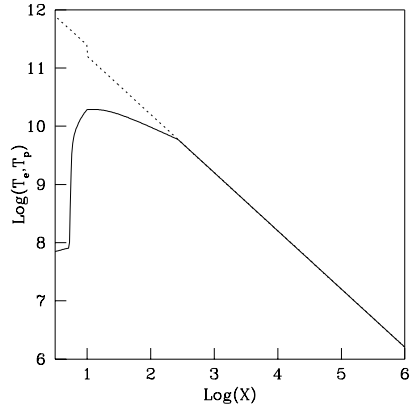}} 
\vskip 0.0cm
\caption{Temperature distribution of protons and electrons as a function of 
radial distance from the black hole.}
\end{figure}

We can vary parameters such as $\Theta$, $X_s$, $R$, $\zeta$ and ${\dot m}=
\frac{\dot M}{\dot M}_{Edd}$, the accretion rate of the flow in units of the
Eddington rate ${\dot M}_{Edd}$. We present a typical  spectrum assuming a
black hole of mass $10M_\odot$. Details are in Mandal \& Chakrabarti (2004).

\begin{figure}[t]
\vskip -2.cm  
\centerline{\epsfxsize=4.1in\epsfbox{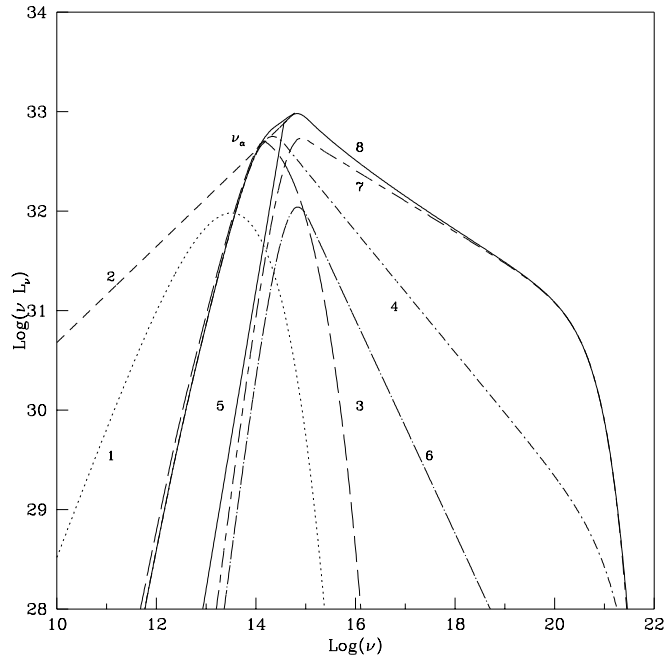}}
\vskip 0.0cm
\caption{A  typical spectrum from an accretion disk is shown with its 
components. See text for details.}
\end{figure}

\section{Results and Interpretations}

Figure 1 shows the variation of the electron and proton temperatures ($T_e$ and
$T_p$ respectively) when ${\dot m}=0.0005$, $R=3.9$, $\Theta=77^o$, $\zeta=0.7$
and $X_s=10$ as a function of the radial distance $X$ (measured in units of 
the Schwarzschild radius $r_g$). The electrons start becoming cooler 
closer to the black hole when the number density gets higher. Higher 
number density increases the cooling for $X\lsim 300$. Very close 
to the black hole, especially after the shock at $X=X_s=10$, 
the splitting is dramatic and electrons become very much cooler.
With our parameters, below $X\sim X_c= 5.8$, the cooling is so strong that 
Comptonization procedure breaks down. We left the temperature to be local 
Keplerian temperature for $X<X_c$. 

In Figure 2, we present the corresponding spectrum with all the contributions
from the accretion flow. Here, different curves are marked 
with a number. The curve marked `1' is due to synchrotron emission from the 
post-shock region (which is also known as the
Centrifugal pressure dominated BOundary Layer  or the CENBOL region)
emitted by Maxwell electrons. The curve marked `2' is that due to
the power-law electrons at CENBOL. The curve marked `3' is the
synchrotron emission from the pre-shock part of the accretion flow. 
The curve marked `4' gives the Comptonized spectra of the synchrotron 
radiation from the pre-shock accretion flow. The curve marked `5' which is
really made up of two pieces, one (up to the self-absorption 
frequency $\nu_a$) of a black-body power-law of slope
$-2$, and the other of a power-law of slope same as curve `2'
due to power-law electrons, gives the net synchrotron emission
from the CENBOL region after the self-absorption is taken care of.
The curve marked `6' is the Comptonization of the soft synchrotron photons 
emitted from Maxwell electrons. The curve marked `7' is the Comptonization of the
soft synchrotron photons emitted from the power-law electrons. 
The curve marked `8' is the total Compton spectra from the pre-shock as well 
as the post-shock regions.

It is clear that the power-law electrons generated at the 
shocks in accretion can leave its signature on the emitted spectrum,
as is evidenced from the power-law part of the spectrum near $Log(\nu)\sim 14-15$.
From $Log(\nu)\sim 16-18$, the power-law photons are produced mainly due to 
Comptonization of the CENBOL photons, while the power-law around 
$Log(\nu)\sim 19-20$ is mainly due to the Comptonization of the photons from 
the pre-shock region. Thus separate regions of the spectrum can 
be identified with separate physical processes inside an accretion disk.

\section{Concluding Remarks}

Our conclusion is that there are several ways a shock may be identified in the
spectrum. At a strong shock, the power-law electrons are produced with a very high
$E_{max}$ and that produces a power-law feature in the spectrum. There is a
bump in the spectrum at around $log_{10}(\nu) \sim 14-15$. On the top of this, 
another bump at around $10^{17}Hz$ would have been present had the Keplerian 
disk been there. 

In Chakrabarti and Titarchuk (1985) soft photons due to a Keplerian disk was
responsible to cool down the post-shock region. There are enough evidence that this
two component advective flow is correct. We used only the sub-Keplerian (or halo) component
in the present paper. Here, considered soft photons to be locally generated due to thermal and magnetic 
bremsstrahlung processes.  In future, we shall incorporate a Keplerian disk
and obtain a combined spectrum. We shall compare these results with 
observed spectra in order to see if these theoretically predicted shocks are
actually present.

\section*{Acknowledgments}
SM acknowledges a RESPOND project from Indian Space Research Organization (ISRO).

\end{document}